\documentclass[twocolumn,showpacs,preprintnumbers,amsmath,amssymb]{revtex4}

\usepackage{graphicx}
\usepackage{rotating}

\begin{document}

\draft

\title{Precision Feshbach spectroscopy of ultracold Cs$_2$}

\author{Cheng Chin\footnote{Present Address: Institut f\"{u}r Experimentalphysik, Universit\"{a}t Innsbruck, Technikerstr. 25, 6020 Innsbruck, Austria}, Vladan Vuleti\'{c}\footnote{Present Address: MIT-Harvard Center for Ultracold Atoms,
 Massachusetts Institute of Technology, Cambridge, MA 02139},  Andrew J. Kerman\footnote{Present Address: Yale University Physics Department, Sloane Laboratory, 260 Whitney Ave., New Haven, CT 06520}, Steven
Chu}
\address{Department of Physics, Stanford University, Stanford, California
94305-4060}

\author{Eite Tiesinga, Paul J. Leo\footnote{Present Address: National Human Genome Research Institute, 49 Convent Drive, Bldg
49, Bethesda, MD 20874-4472 }, and Carl J. Williams}
\address{Atomic Physics Division, National Institute of Standards
and Technology, Gaithersburg, Maryland 20899-8423}

\begin{abstract}
We have observed and located more than 60 magnetic field-induced
Feshbach resonances in ultracold collisions of ground-state
$^{133}$Cs atoms. These resonances are associated with molecular
states with up to four units of rotational angular momentum, and
are detected through variations in the elastic, inelastic, and
radiative collision cross sections. These observations allow us to
greatly improve upon the interaction potentials between two cesium
atoms and to reproduce the positions of most resonances to
accuracies better than $0.5\%$. Based on the relevant coupling
scheme between the electron spin, nuclear spin, and orbital
angular momenta of the nuclei, quantum numbers and energy
structure of the molecular states beneath the dissociation
continuum are revealed. Finally, we predict the relevant collision
properties for cesium Bose-Einstein condensation experiments.
\end{abstract}

\pacs{34.50.-s, 05.30.Jp, 32.80.Pj, 67.40.Hf}

\maketitle

\narrowtext

\section{Introduction}

The collision properties of an ultracold and dilute atom gas are
strongly influenced by the long-range interactions between two
atoms. When the interaction potential supports a weakly bound
state near the scattering energy, the atomic collision properties
can be resonantly altered, a situation referred to as a Feshbach
resonance \cite{fesh0, fesh}. In many cold-atom systems,
magnetically tunable Feshbach resonances have been discovered, and
have led to ground-breaking observations including the implosion
of a Bose-Einstein condensate (BEC) \cite{implo}, the coherent
coupling between an atomic BEC and molecules \cite{cohr}, the
creation of bright solitons \cite{sol}, and recently the creation
of ultracold molecules \cite{chi03, mol, csmol} and of a molecular
BEC \cite{molbec}.

The collisional properties of ultracold cesium atoms have
intrigued experimentalists and theorists because of their large
clock shifts \cite{gib93}, enormous collision cross sections
\cite{xsec}, and extreme difficulty to reach BEC \cite{becfail}.
These anomalies in the atom-atom scattering can be explained by
the coupling of the scattering continuum to molecular states.
While these states cannot be accessed by conventional spectroscopy
\cite{wei85}, they may be tuned into resonance with the scattering
continuum and induce Feshbach resonances. Detection of multiple
Feshbach resonances, or Feshbach spectroscopy, permits a precise
determination of the long-range interaction parameters, as well as
the molecular structure near threshold. With this information, the
cold collision anomalies can be resolved and the clock shifts
\cite{clock}, collision cross sections and scattering lengths can
be accurately calculated  \cite{leo00}.

In this work, we report the observation of more than 60 Feshbach
resonances of cesium atoms in 10 different incident channels. In
particular, we employ a radiative detection scheme to resolve
narrow resonances \cite{chi03, vul99} whose locations allow us to
significantly improve our determination of Cs interaction
parameters over our previous work \cite{chi00, leo00}. With these
parameters, the molecular energy structure near threshold, as well
as $s$-wave scattering lengths and collision properties can be
precisely determined.

We organize the paper as follows. First, we outline the
experimental setup and procedures, and the general measurement
methods in Sec.~\ref{sec:exp}. We present and discuss the results
from inelastic Feshbach spectroscopy, elastic Feshbach
spectroscopy and radiative spectroscopy in Sections
\ref{sec:inelast}, \ref{sec:elast}, and \ref{sec:rad},
respectively.  In Section \ref{sec:theory} we discuss the
Hamiltonian for two ground-state cesium atoms and approximate
quantum numbers in the system.  The numerical procedures for
calculating scattering properties and bound state energies are
presented in Sec.~\ref{sec:Num}.  Section \ref{sec:results}
analyses all observed Feshbach resonances and assigns quantum
numbers to each resonance.

\section{Experiment} \label{sec:exp}
\subsection{Feshbach spectroscopy}
\begin{figure}
\includegraphics[width=3in]{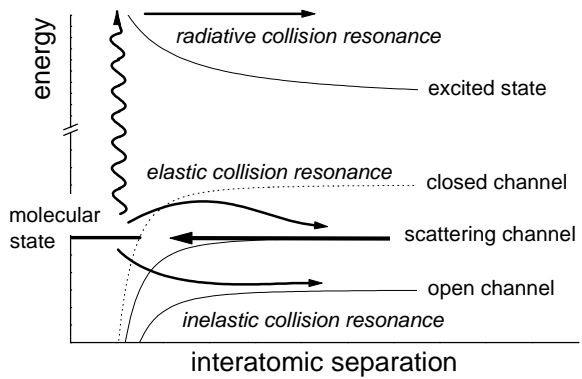}
\caption{Illustration of radiative, elastic, and inelastic
Feshbach resonance. The molecular state is supported by the closed
channel.} \label{processes}
\end{figure}

Feshbach resonance in binary atomic collisions is illustrated in
Fig.~\ref{processes}. Interacting atom pairs in the scattering
continuum, or scattering channel, couple to a discrete bound state
supported by a closed channel with higher internal energy. This
coupling resonantly alters the outgoing scattering amplitude in
the scattering channel (elastic collision resonance) and in the
lower-lying open channels (inelastic collision resonance).

In ultracold atom experiments, Feshbach resonances can be observed
in elastic and inelastic collision rates \cite{fesh, Vul99} and in
the molecular population, probed by photoassociation
\cite{Heinzen}, or photodissociation \cite{chi03} as shown in
Fig.~\ref{processes}.

In most cases the magnetic moments of the molecular state and the
scattering continuum are unequal, such that the energy difference
between the molecular state and the colliding atom pair can be
tuned by means of an external magnetic field. The molecular
bound-state energy structure near dissociation threshold is then
reflected in the magnetic-field dependence of the atom-atom
scattering. A measurement of the ultracold collision properties as
a function of magnetic field (Feshbach spectrum) in combination
with theoretical modeling thus unveils the underlying molecular
spectrum. In this work, we are able to associate each resonance
with its quantum numbers of both the incoming scattering channel
and the molecular state, where it is noteworthy that the $s$-,
$p$-, $d$- \dots partial wave of the incoming channel and of the
resonance need not be the same. A collection of Feshbach spectra
for colliding atoms in several different quantum states then
provides the information necessary to accurately determine the
long-range interaction parameters and the molecular energy
structure near the dissociation continuum.

\subsection{Preparation of ultracold high-density samples}

In order to measure Feshbach spectra with good signal-to-noise
ratio and high resolution, and to simplify the theoretical
analysis, it is favorable to use high-density samples of ultracold
atoms confined in a trap that produces negligible tensor light
shifts of the atomic and molecular energy levels \cite{deu98}. We
use a far-detuned, linearly polarized, one-dimensional ($1D$)
optical lattice trap to confine the atoms, in conjunction with an
optical cooling method that can produce high-density samples at
$\mu$K temperatures \cite{vul98,ker00}.

\begin{figure}
\includegraphics[width=2in]{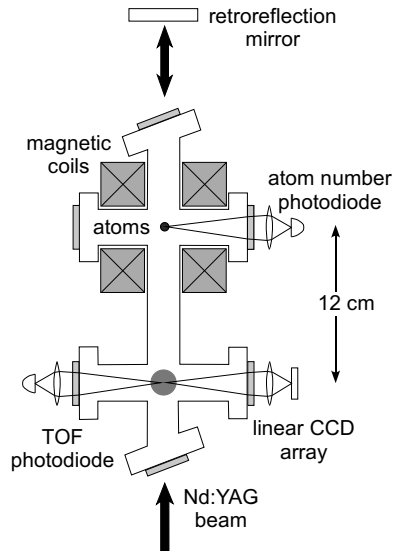}
\caption{Apparatus for Feshbach spectroscopy of $^{133}$Cs.}
\label{setup}
\end{figure}

Fig.~\ref{setup} shows schematically the experimental setup.
Ultracold samples of cesium atoms are prepared in the upper part
of a compact ultrahigh vacuum chamber. A pair of magnetic field
coils, when operated in the anti-Helmholtz configuration with
opposite currents, provides the spherical quadrupole field for a
magneto-optical trap (MOT), and, in the Helmholtz configuration,
produces a homogeneous magnetic field up to 25 mT. A linearly
polarized, vertically propagating, retroreflected Nd:YAG laser
beam provides the far-detuned one-dimensional  lattice dipole trap
at 1064 nm \cite{vul98}. The number of trapped atoms is inferred
from the fluorescence emitted by the cloud when it is illuminated
with resonant light on the $F_g=4 \rightarrow F_e=5$ hyperfine
component of the $D_2$ line near 852 nm. Cloud temperatures in the
vertical and horizontal directions are determined by releasing the
atoms and performing time-of-flight imaging 12 cm below onto a
photodiode and a linear CCD array, respectively. Trap vibration
frequencies are measured by parametric excitation \cite{fri98}.
The atomic density is derived from the measured atom number,
temperature and trap vibration frequencies, as detailed in Ref.
\cite{chi00}.

The ultracold sample is prepared by first collecting $5 \times
10^8$ atoms in a vapor-cell magneto-optical trap (MOT) in $500$
ms. Superimposed with the MOT is the YAG $1D$ optical lattice
dipole trap that at a power of 8 W and a beam waist of 260 $\mu$m
provides a trap depth of $U/h = 1.6$ MHz, and axial and radial
vibration frequencies of $\omega_{a}/2\pi =50$ kHz and
$\omega_{r}/2\pi =80$ Hz, respectively. The preparation of the
high-density sample in the YAG dipole trap is accomplished by
means of two phases of Raman-sideband cooling (RSC) with a $3D$
near-detuned optical lattice \cite{ker00}. We extinguish the MOT
light and apply the first RSC for 10 ms in a small bias field of 5
$\mu$T, which cools the atoms to temperatures below 1 $\mu$K, and
predominantly polarizes them into the lowest-energy magnetic
sublevel $|6S_{1/2},F=3,m_F=3\rangle$. Here $F$ is the total
angular momentum of cesium atoms in the $6S_{1/2}$ ground state
and $m_F$ is the angular momentum projection along the magnetic
field direction. When the $3D$ near-detuned optical lattice is
extinguished in 1 ms, the release of the atoms into the $1D$ YAG
lattice trap is adiabatic only in the vertical direction, and we
observe a radial oscillation as the atoms slide down the trap
potential. To remove some of the excess potential energy, we wait
for 4 ms until the atoms have the greatest kinetic energy and
highest density, and perform a second phase of RSC \cite{dep01}.
Due to the high density near $10^{12}$ cm$^{-3}$, the second
cooling requires a weaker optical pumping intensity and longer
cooling time of 15 ms. After rethermalization for 200 ms in the
YAG lattice trap, $1\times 10^8$ atoms are prepared at a
temperature of 3 $\mu$K to 5 $\mu$K with a vertical (horizontal)
rms radius of $\sigma_z =580$ $\mu$m ($\sigma_x =30$ $\mu$m). This
means that near the central region, each site in the 1D YAG
lattice trap contains $4\times 10^4$ atoms at a mean density of
approximately $1 \times 10^{13}$cm$^{-3}$. The two phases of RSC
also allow us to adjust almost independently the atomic density
and temperature by changing the detuning of the optical pumping
beam during the first and the second RSC phase, respectively.

The preparation of pure samples in desired target states is
crucial for cold collision experiments because different internal
states can have drastically different collision properties. To
improve the atomic polarization over that achieved by the RSC
alone, we apply an additional optical pumping pulse for 3 ms
during the release from the second RSC phase. This has the
advantage that atom-atom collisions are still suppressed due to
the $3D$ confinement in the near-detuned optical lattice, which
prevents radiative collision loss. Using microwave spectroscopy,
we optimize the optical pumping process at a bias field of 14
$\mu$T, and prepare up to $98\%$ of the atoms in one of the
stretched states $|F,m_F=\pm F\rangle$, and the remaining atoms in
the neighboring $|F,m_F=\pm (F-1)\rangle$ state, where $F$=3 or 4.
Feshbach resonances in four scattering states or channels can be
thus studied: $(3,3)+(3,3)$, $(3,-3)+(3,-3)$, $(4,4)+(4,4)$ and
$(4,-4)+(4,-4)$, where $(F_1,m_{F1})+(F_2,m_{F2})$ indicates the
collision of one atom in the $|F_1,m_{F1}\rangle$ state with one
in the $|F_2,m_{F2}\rangle$ state.

We also prepare mixed samples containing atoms in two different
internal states by either detuning the optical pumping beam, or by
applying an additional microwave pulse to transfer part of the
population. Typically, we prepare $90\%$ of the population in a
stretched state $|a\rangle$, and $10\%$ in another state
$|b\rangle$. After fully characterizing the collision properties
of the $(a)+(a)$ channel, the mixed collisions $(a)+(b)$ can be
monitored by either identifying resonances growing with the
population in $|b\rangle$, or by selectively detecting only the
population in $|b\rangle$. The large population ratio of the two
states ensures that $(b)+(b)$ collision processes remain
insignificant. Mixed scattering channels investigated in this work
include $(3,3)+(3,2)$, $(3,3)+(4,4)$, $(3,3)+(4,3)$, $(3,3)
+(4,2)$,$(3,-3)+(3,-2)$ and $(4,-4)+(4,-3)$.

The ability to study all these different scattering channels is
crucial to provide information on the molecular bound states near
the $(F_1=3)+(F_2=3)$, $(F_1=3)+(F_2=4)$, and $(F_1=4 )+(F_2=4)$
hyperfine asymptotes. These span roughly h 18 GHz in binding
energy. As was shown in Ref.~\cite{leo00}, the molecules have
different characteristics near the three dissociation limits,
which particularly helps determine the strength of the van der
Waals interaction.

After the atoms have thermalized in the dipole trap, a uniform
magnetic field up to 25 mT is applied. In order to preserve the
atomic polarization during the field ramp, we first increase the
field from 14 $\mu$T to 200 $\mu$T in 200 ms, and then to an
arbitrary field value in another 100 ms. The magnetic field
experienced by the atoms is calibrated to an accuracy of $\delta
B<0.1\mu T$ at low field and $\delta B/B <10^{-4}$ at high field
from the Zeeman splitting between magnetic sublevels, as measured
with microwave spectroscopy \cite{chi00}. Slowly changing stray
fields of typically 50 $\mu$T and remnant fields from the
magnetized vacuum chamber up to 2 $\mu$T are carefully cancelled
with six independent bias coils in three orthogonal directions to
an accuracy of 0.1 $\mu$T for the field, and 2 $\mu$T/cm for the
field gradient. An effective magnetic field due to residual
circular polarization of the YAG trapping beam \cite{deu98} is
monitored with microwave spectroscopy, and reduced below 0.05
$\mu$T by linearizing the beam polarization. To suppress the
effects of field inhomogeneity and atomic density variation
between the various YAG 1D lattice sites, we perform measurements
only on the center portion of the cloud within $\pm 0.4\sigma_z$,
which contains 900 lattice sites with a mean atomic density
variation of $<10\%$.

\section{Inelastic Feshbach spectroscopy} \label{sec:inelast}

Inelastic collisions occur when the initial scattering state
couples to open channels with lower internal energy, see
Fig.~\ref{processes}. Due to energy conservation, the internal
energy difference is converted into relative kinetic energy of the
atom pair, and is either on the order of the ground state
hyperfine splitting or the Zeeman energy.  In both cases, the
energy release is generally much larger than our trap depth and
therefore results in a loss of the colliding atom pair from the
trap. Near a Feshbach resonance, inelastic loss is either enhanced
or suppressed due to constructive or destructive interference
between the off-resonant scattering amplitude with the on-resonant
amplitude. In this work, we observe mostly enhanced inelastic
collision processes, and only one prominent suppression at $7.66$
mT in the $(3,3)+(4,2)$ incident scattering channel \cite{chi00}.

\begin{figure}
\includegraphics[width=3in]{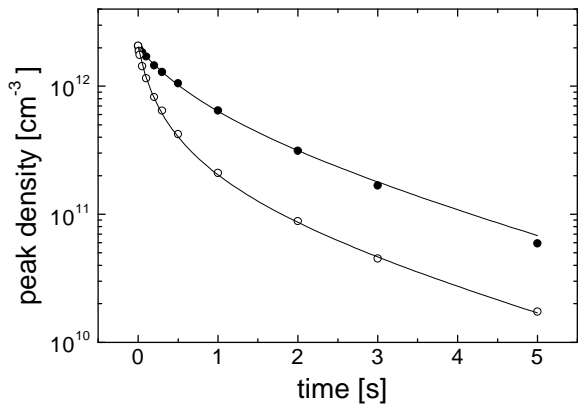}
\caption{Resonant and off-resonant time evolution of the atomic
density. Atoms are prepared in $|F=4,m_F=-4 \rangle$ at $T = 5.3$
$\mu$K and $\bar{n}= 2.1 \times 10^{12}$ cm$^{-3}$ at $B=14$
$\mu$T (solid circles) and $B=20.53$ mT (open circles), where a
strong Feshbach resonance is located. See also Table
\ref{t:el_inel_1}.} \label{fig1}
\end{figure}

Experimentally, the inelastic rates are determined by fitting the
mean atom density, $\bar{n}$, to $d\bar{n}/dt=
-L\bar{n}-K\bar{n}^2$, where $L$ is an one-body loss rate, and $K$
is the desired collision rate coefficient.  If we assume that the
atoms only reside in the harmonic region of the YAG dipole trap
potential and that the cloud is in thermal equilibrium and at
constant temperature $T$ during the experiment, then the mean atom
density is $\bar{n}=N(m\bar{\omega}^2/4\pi k_BT)^{3/2}$ where $N$
is the atom number in one lattice site,
$\bar{\omega}=\omega_z^{1/3}\omega_r^{2/3}$ is the mean vibration
frequency, $m$ is the cesium atomic mass, and $k_B$ is the
Boltzmann constant.  The above differential equation can be solved
analytically and directly relates the measured trap loss to the
thermally averaged loss coefficient $K$. The one-body loss rate is
the same for all measurements. A typical measurement is shown in
Fig.~\ref{fig1} with a holding time up to $5$ s.

The loss rates as a function of external magnetic field are
measured by observing the atom loss within a holding time between
$30$ ms to $300$ ms, such that the maximum collision loss is less
than $30\%$, and the atomic temperature varies by less than $10\%$.
Slow variations in the initial atom number are monitored after every $10$
measurements, and corrected for. On the other hand, to observe and locate weak
resonances, we allow the atoms to interact for a longer time up to
$500$ ms in order to obtain better signal to noise. In this case,
we ignore the temperature evolution and report only the fractional
loss of atoms.

Collisions between atom pairs in two different internal states,
such as $(3,-3)+(3,-2)$, differ from those in the same state in
that both the even and odd partial waves can be scattered. In the
$(3,-3)+(3,-2)$ channel, for instance, we observe rich Feshbach
spectra containing both $s$- and $p$-wave resonances.

Positions of the inelastic Feshbach resonances are tabulated in
Tables \ref{t:el_inel_1} and \ref{t:el_inel_2}.

\subsection{(4,4)+(4,4) inelastic collisions}

The stretched $|4,4\rangle$ state was considered a promising
candidate for reaching Bose-Einstein condensation in a magnetic
trap before the cesium collision properties were revealed. Large
inelastic losses that grow as the temperature is reduced were
discovered that prevent the condensation of cesium in this state
\cite{arn97,sod98,gue98}.

We find no resonance structure for a sample polarized in the
$|4,4\rangle$ state at 5$\mu$K, but a loss rate coefficient
increasing slowly from $2\times 10^{-12}$ cm$^3$s$^{-1}$ to
$3\times 10^{-12}$ cm$^3$s$^{-1}$ over the range of $B=0$ to $25$
mT. The absence of resonances in this channel is expected since
the scattering channel $(4,4)+(4,4)$ has the highest hyperfine and
Zeeman energy of all ground-state hyperfine levels, and
consequently Feshbach resonances cannot occur.

\subsection{(4,-4)+(4,-4) and (4,-4)+(4,-3) inelastic collisions}

We observe two narrow inelastic Feshbach resonances for collisions
between two $|4,-4\rangle$ atoms: a weak one at $10.59$ mT and a
strong one at $20.50$ mT. In addition we have discovered an
inelastic resonance in the $(4,-4)+(4,-3)$ channel at $20.66$ mT.
The resonance field values have a ratio of $1:2:2$ to within
$3\%$.  This ratio is not a coincidence, but a result of the
molecular bound state structure below the continuum. See Table
\ref{t:el_inel_1}. All three resonances are identified as
originating from bound states with identical binding energy at
zero magnetic field.

\subsection{(3,-3)+(3,-3) and (3,-3)+(3,-2) inelastic collisions}

\begin{figure}
\includegraphics[width=3.2in]{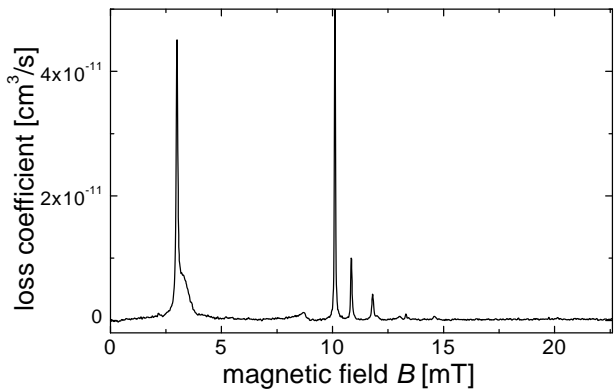}
\caption{Binary loss coefficient in a gas of $|F=3,m_F=-3\rangle$
atoms as a function of external magnetic field. The initial mean
atomic density is $\bar{n}=5\times 10^{12}$ cm$^{-3}$ at a
temperature of $\sim5$ $\mu$K. Populations in the
$|F=3,m_F=-3\rangle$, $|F=3,m_F=-2\rangle$ and all other states
are $95\%$, $\sim 5\%$ and $<1\%$, respectively. Magnetic field
resolution is $10\mu$T.} \label{fig3}
\end{figure}

The $|3,-3\rangle$ state can be magnetically trapped and several
attempts to reach BEC in this state were thwarted by the large
collision loss and negative scattering length at low magnetic
field \cite{gue98}. In this work, we observe multiple inelastic
resonances for collisions between two such atoms, as shown in
Fig.~\ref{fig3}. One resonance at 2.18 mT is not identified, see
Table \ref{t:el_inel_2}.

\begin{figure}
\includegraphics[width=3in]{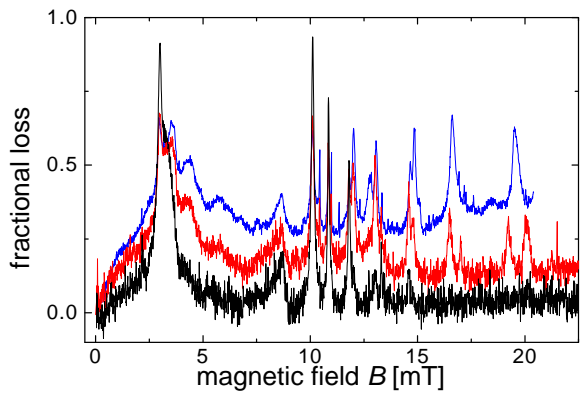}
\caption{Atom loss as a function of magnetic field for an initial
mean density $\bar{n}=5\times 10^{12}$cm$^{-3}$, temperature of
$\sim 5$ $\mu$K and interaction time of $300$ ms. Populations in
the $|F=3,m_F=-3\rangle$, $|F=3,m_F=-2\rangle$ and all other
states are $95\%$, $\sim5\%$ and $<1\%$ for the lower curve,
$85\%$, $\sim10\%$ and $\sim 5\%$ for the middle curve and $\sim
70\%$, $\sim20\%$ and $\sim 10\%$ for the upper curve,
respectively. Comparison between the curves facilitates the
identification of the incident channel responsible for the
Feshbach resonances. Magnetic field resolution is $10\mu$T.}
\label{fig4}
\end{figure}

In $(3,-3)+(3,-2)$ collisions, multiple $s$- and $p$- wave
resonances are found and identified by varying the population in
the $|3,-2\rangle$ state, shown in Fig.~\ref{fig4}. An alternative
method to identify mixed state resonances is based on detecting
the population in $|3,-2\rangle$ by microwave transitions on
samples with $90\%$ of the population in $|3,-3\rangle$ and $\sim
10\%$ in $|3,-2\rangle$. In this case, we observe enhanced loss of
the $|3,-2\rangle$ population near inelastic collision resonances
in the $(3,-3)+(3,-2)$ channel, and suppressed loss of the
$|3,-2\rangle$ population near inelastic collision resonances in
the $(3,-3)+(3,-3)$ channel, the latter being due to the reduction
in density of $\vert 3,-3 \rangle$ atoms. This provides a clear
distinction between the two collision processes.

\subsection{(3,3)+(3,3) and (3,3)+(3,2) inelastic collisions}

The $(3,3)+(3,3)$ channel is the lowest hyperfine scattering
channel, and therefore has no binary exothermic collision
processes. Collisional inelastic loss is then only due to
three-body recombination, a process that falls outside the scope
of this paper. A quantitative study of the recombination loss in
the $|3,3\rangle$ state is given in Ref.~\cite{tino03}. In
collisions between $|3,3\rangle$ and $|3,2\rangle$ atoms we have
observed no resonances below $B$=23.5 mT. Theory predicts weak
resonances that are beyond the sensitivity of our current
experiment.

\subsection{(3,3)+(4,2), (3,3)+(4,3), and (3,3)+(4,4) inelastic collisions}

Multiple Feshbach resonances due to $(3,3)+(4,2)$ and
$(3,3)+(4,3)$ scattering are observed in samples with $80\%$ population
in $|3,3\rangle$, and $15\%$ in $|4,3\rangle$ or $|4,2\rangle$. The
remaining atoms are predominantly in the $|3,2\rangle$ hyperfine
state. We have verified that $(3,3)+(3,2)$, $(4,3)+(3,2)$ and
$(4,2)+(3,2)$ processes do not contribute to the loss.

We prepare the sample by first polarizing $95\%$ of the atoms in
the $|3,3\rangle$ state at  $B=14\mu$T, and then apply a microwave
pulse for a few ms, which selectively transfers approximately
$15\%$ of the population into $|4,3\rangle$ or $|4,2\rangle$. Due
to the almost identical energy splitting between the $|3,3\rangle
\leftrightarrow |4,2\rangle$ and $|3,2\rangle \leftrightarrow
|4,3\rangle$ microwave transitions at low field, we have a small
population of $|4,3\rangle$ in the experiment aimed at finding
resonances in a $(3,3)+(4,2)$ collision. Pollution by
$|3,3\rangle+|4,3\rangle$ resonances occurs, as was previously
reported in Ref.~\cite{chi00}.

We also observe a weak pollution by $(3,3)+(4,2)$ resonances in
the $(3,3)+(4,3)$ spectrum. This cannot be explained by microwave
transitions, since $|3,3\rangle \leftrightarrow |4,3\rangle$ and
$|3,2\rangle \leftrightarrow |4,2\rangle$ do not have the same
frequency. A possible process that creates atoms in the
$|4,2\rangle$ state is the inelastic collision process $
(3,2)+(4,3) \rightarrow (3,3)+(4,2) $. At low fields the hyperfine
and Zeeman energy difference between the final and initial state
is $\delta E=k_B 0.52\mu$K $\times (B/mT)$. For our atomic
temperatures and magnetic fields, this endothermal spin-changing
collisions can create atoms in the $|4,2\rangle$ state.

The four resonances in the $(3,3)+(4,2)$ channel at $6.17$ mT,
$8.38$ mT, $11.0$ mT and $11.2$ mT, and all four resonances
observed in the $(3,3)+(4,3)$ channel at $12.9$ mT, $17.3$ mT,
$22.7$ mT and $23.0$ mT are paired with identical ratio of the
field values $2.07(2)$.  This observation is confirmed by the
theoretical identification of the paired resonances as being due
to molecular states with identical binding energy at zero magnetic
field.

In a separate experiment we have found no inelastic resonances in
$(3,3)+(4,4)$ collisions for magnetic field strengths up to 23.5
mT, in agreement with theory.

\section{Elastic Feshbach spectroscopy} \label{sec:elast}

The resonant change of the scattering amplitude in the incident
channel also results in a modification of the elastic cross
section. We refer to this process as an elastic Feshbach
resonance, see Fig.~\ref{processes}. Beside the direct measurement
of the thermalization rate between the axial and radial motion
\cite{rob98,Vul99}, we have developed a more sensitive measurement
technique for elastic Feshbach resonances, the measurement of the
evaporation rate in a shallow trap \cite{chi00}. This method
converts the temperature evolution measurement into an atom number
measurement, which provides better sensitivity and signal-to-noise
ratio.

We report elastic collision properties in the pure $(3,3)+(3,3)$
channel and in the mixed $(3,3)+(3,2)$ channel, where the
measurements are not complicated by inelastic processes. The
elastic spectrum of the $(3,3)+(3,3)$ channel has previously been
reported in Refs. \cite{Vul99,chi00}. Although the elastic cross
section can in principle be both enhanced and decreased by the
Feshbach resonance, we observe only dips in the elastic cross
section at $T$=5 $\mu$K. This is due to the large background
scattering length, resulting in an elastic cross section that is
mostly unitarity limited to $\sigma = 8\pi/k^2$ ($\sigma =
4\pi/k^2$) when the colliding cesium atoms are (are not) in the
same state. A vanishing elastic collision cross section at $1.7$
mT resulting from a strong $s-$wave Feshbach resonance in the
$(3,-3)+(3,-3)$ channel was first observed as a reduction in the
cross axis thermalization \cite{Vul99}. Three additional $d-$wave
Feshbach resonances were observed by monitoring the evaporative
loss rate \cite{chi00}. A vanishing elastic cross section in the
$(3,3)+(3,2)$ channel is observed with the same evaporative loss
method, where we measure the remaining atom number in $|3,2
\rangle$. Positions of the elastic collision rate minima are given
in Table \ref{t:el_inel_1}.

The elastic collision measurements based on evaporative loss are
limited to resonances whose width in energy is comparable to or
exceeds the thermal energy of the sample $k_B T$. For narrow
resonances the resonant variation of the elastic collision rate
over a small interval of collision energies does not lead to
observable variations in the evaporation rate. For the detection
of such weak resonances we have to resort to radiative Feshbach
spectroscopy, discussed in the next section.

\section{Radiative Feshbach spectroscopy}\label{sec:rad}

Narrow resonances with a linewidth that is small compared to the
thermal energy $k_B T$ do not significantly affect the sample's
thermalization rate, and therefore cannot be probed by elastic
Feshbach spectroscopy. However, it is possible to directly measure
the enhanced (quasi-)bound state population on a Feshbach
resonance using a far-detuned laser beam to selectively dissociate
the molecules, while leaving the atoms unperturbed. This technique
is called radiative Feshbach spectroscopy. \cite{chi03, vul99}.

In this work, we apply radiative Feshbach spectroscopy to probe
Feshbach resonances in the $(3,3)+(3,3)$ and $(3,3)+(3,2)$
scattering channels using a probe beam typically detuned 20 GHz to
4 THz to the blue of the cesium $D_2$ transition at 852.3 nm. The
probe beam is provided by a titanium-sapphire laser, and uniformly
illuminates the atom sample with a stabilized intensity up to
$50$W/cm$^2$. The intensity and detuning of the laser are adjusted
such that the single-atom excitation is sufficiently weak, and negligible
atom loss is observed when the magnetic field is tuned off the
Feshbach resonances, while maximizing the loss on the Feshbach
resonance.

\begin{figure}
\includegraphics[width=3in]{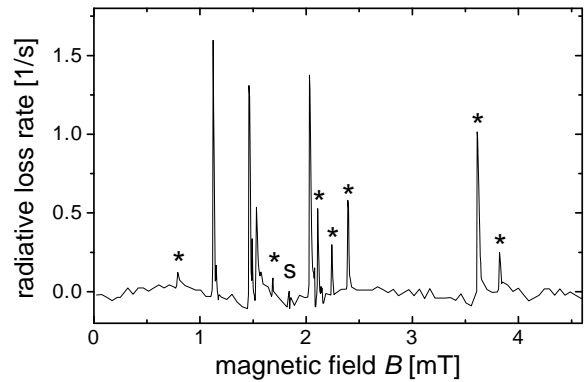}
\caption{A detailed radiative loss spectrum with a field
resolution of $0.5 \mu$T. The probe beam at wavelength $\lambda =
844$ nm has an intensity of $50$W/cm$^2$. $85\%$ ($15\%$) of the
atoms are in the $|3,3\rangle$ ($|3,2\rangle$) state. The
temperature and mean density of the sample are $3.5\mu$K and
$1\times 10^{13}$ cm$^{-3}$, respectively. ``S" indicates a shape
resonance; the stars indicate Feshbach resonances in the
$(3,3)+(3,2)$ channel. Magnetic field resolution near the
resonances is $5\mu$T.} \label{fig5}
\end{figure}

To measure the radiative collision loss, we first prepare atom
samples either fully polarized in the $|3,3\rangle$ to study
$(3,3)+(3,3)$ collisions, or $85\%$ in $|3,3\rangle$ and $15\%$ in
$|3,2\rangle$ to study $(3,3)+(3,2)$ collisions. After the
preparation and field ramping, we typically illuminate the atoms
with the probe beam for t=100 ms to 300 ms, and measure the trap
loss. To determine the radiative loss rate as a function of
magnetic field, we perform two consecutive atom number
measurements at every magnetic field value $B=B_i$ to discriminate
non-radiative loss: $N_1$ is the atom number at $B=B_i$ without
the probe beam, and $N_2$ is obtained at $B=B_i$ with the probe
beam. The radiative loss rate is then given by
$\gamma=(1-N_2/N_1)/t$, where the illumination time $t$ is chosen
to keep the maximum atomic loss below $30\%$. To increase
detection sensitivity on weak resonances, we lengthen $t$ to $>
500$ms, and average each data point up to 5 times. A detailed
spectrum thus obtained is shown in Fig.~\ref{fig5}. The details of
the radiative loss line shape are discussed in Ref. \cite{vul99};
atom-molecule dynamics and the sensitivity of the our radiative
Feshbach spectroscopy are studied in Ref. \cite{chi03}. Positions
of the resonances are tabulated in Table \ref{t:radiative}. One of
the Feshbach resonances observed by radiative Feshbach
spectroscopy was recently used to create ultracold cesium
molecules \cite{csmol}.

\section{Theory}\label{sec:theory}

The structure of the Hamiltonian of two interacting $^2S$ ground
state alkali-metal atoms is well known.  It contains the atomic
kinetic energy operator, an atomic Hamiltonian for each atom, two
Born-Oppenheimer potentials with symmetry $^1\Sigma_g^+$ and
$^3\Sigma_u^+$, the nuclear rotation operator
$\hbar^2\vec{l}^2/(2\mu R^2)$, and weaker relativistic spin-spin
dipole and second order spin-orbit interactions.  Here $\vec{l}$
is the nuclear mechanical angular momentum and $\mu$ is the
reduced mass of the molecule.

The atomic Hamiltonian contains a Fermi contact term and the
Zeeman interaction when an external magnetic field, $B$, directed
along the $z$ axis is present during the collision.  The
eigenstates of the $B=0$ atomic Hamiltonian are
$|F_\alpha,m_{F\alpha}\rangle$, where $\alpha=1$ or $2$ for atom
$1$ or $2$, $\vec{F}_\alpha=\vec{s}_\alpha +\vec{i}_\alpha$ and
$m_{F\alpha}$ is the projection of $\vec{F}_\alpha$ along the $z$
axis. Here $\vec{s}_\alpha$ and $\vec{i}_\alpha$ are the atomic
electron and nuclear spin, respectively.  For $B>0$, states with
the {\it same} $m_{F\alpha}$ mix, $F_\alpha$ is no longer a good
quantum number, and the zero field $m_{F\alpha}$ degeneracy is
lifted. For convenience, atomic eigenstates in a magnetic field
will be labeled by $|F_\alpha,m_{F\alpha}\rangle$ since for fields
used in this paper the Zeeman interaction is small compared to the
hyperfine interaction and $F_\alpha$ is an approximately good
quantum number. We use hyperfine constants and magnetic moments
from Ref.~\cite{Ari77}.  The nuclear spin of cesium is $7/2$.

The selection rules for the Born-Oppenheimer Hamiltonian conserve
$\vec{l}$ and $\vec{f}=\vec{F}_1+\vec{F}_2$.  Consequently, bound
states and scattering amplitudes can be labeled by $fl$ for $B=0$
and $m_fl$ for $B>0$. Here, $m_f=m_{F1}+m_{F2}$. The two
relativistic interactions weakly mix states with different $l$ and
$f$.  Global symmetries ensure that the molecular Hamiltonian,
including the relativistic interactions, conserves parity and
total angular momentum $\vec{F}=\vec{f}+\vec{l}$ and its
projection $M=m_f+m_l$ for $B=0$ or only $M$ for nonzero fields.
Only even or odd partial waves are coupled.  For $B=0$ there are
at most 72 coupled channels while for non-zero field there are
infinitely many coupled channels. In practice the number of
channels is restricted by using knowledge about the relative
strengths of the individual terms in the Hamiltonian.

The relativistic interactions, even though weak, are crucial in
understanding the presence of $g$-wave Feshbach resonances in our
gas of ultra-cold Cs atoms.  The temperature $T\approx$ 5 $\mu$K
is small compared to the 200 $\mu$K and 1 mK barrier height of the
$d$- and $g$-partial waves, respectively, and only incoming
$s$-wave collisions contribute to the experimental signal.
Consequently, coupling from $s$- to $d$- and $g$-waves and back to
$s$-waves is necessary.  Small corrections to the location of the
Feshbach resonances due to relativistic interactions can be
observed.

\section{Numerical approaches} \label{sec:Num}

The scattering properties and bound-state energies of the ground
state Hamiltonian are obtained with two separate numerical
approaches. For scattering the wavefunction at energy $E$ is found
using a Gordon propagator\cite{Gordon}.  From the scattering
wavefunctions elastic and inelastic rate coefficients as a
function of magnetic field can be evaluated. Feshbach resonances
appear as sharp peaks or dips in the magnetic field dependence of
the rate coefficients. For comparison with the experiments the
rate coefficients need to be thermally averaged.

Obtaining discrete bound states with the Gordon method is
cumbersome since eigenenergies are not {\it a priori} known and
multi-channel scattering wavefunctions need to be calculated at a
large number of energies $E$. Consequently, a discrete variable
representation~\cite{DVR} for the radial kinetic energy operator
is used to find the bound states. In this approach the eigenvalues
of a linear system of size given by the number of radial
collocation points times the number of coupled channels need to be
calculated. This can be done with standard linear algebra
packages.  However, resource limitations tend to restrict the
number of coupled channels that can be conveniently handled.  For
the heavy cesium dimer a realistic maximum number of channels lies
between 10 to 15, although 20 channels can still be treated.  An
interesting alternative for finding bound state energies, which
does not require the storage of the linear system, can be based on
the multi-channel quantum defect theory~\cite{MQDT}. It takes
advantage of analytic properties of wavefunctions as a function of
energy $E$ in order to limit the number of wavefunction
evaluations.

Bound states are calculated over a range of magnetic field values.
Feshbach resonances occur when a bound state crosses a collisional
threshold.  Examples of resonances can be found in
Refs.~\cite{Vog98,Laue02}. In the absence of the two relativistic
interactions calculations label individual bound state by $m_f$,
$l$, and $M$.  In addition, the field dependence of a level can be
traced to a bound state at zero field. At zero field $f$ labels
the bound states. For magnetic fields used in our experiment $f$
remains approximately good. Coupling of states with different
$m_f$ but the same $l$ are sometimes needed to fully assign the
Feshbach resonances. For this paper bound state calculations are
used to assign quantum numbers to the resonances.

The Feshbach resonances are experimentally observed in either
elastic cross sections, inelastic rate coefficients, or radiative
collision rates. The former two measurements can be modeled from
first-principle scattering calculations of cross sections and rate
coefficients.  In principle the rate coefficients need to be
thermally averaged.  However, a proper thermalization was
impractical and for a comparison between theory and experiment
only a single collision energy given by the mean collision energy
of a gas at temperature $T$ was used.  A combination of narrow
Feshbach resonances and the need to study the effect of variations
in the shape of the two Born-Oppenheimer potentials on the
resonance locations would lead to an untenable number of
scattering calculations.  A one-standard-deviation uncertainty of
0.02 mT to 0.05 mT in the calculated magnetic-field location of
the Feshbach resonances observed in elastic or inelastic rates
results from the use of a single collision energy.

The radiative collision data require modeling of the rate
coefficient for the absorptions of a photon by a pair of
ultra-cold Cs atoms \cite{Nap94}. However, we are interested in
reproducing the location of Feshbach resonances, and not in the
absolute absorption rates. Consequently, from a theoretical
perspective it is sufficient to locate the resonances in the
elastic cross section in the absence of light. The $l=4$
($g$-wave) resonances observed by radiative Feshbach spectroscopy
could not be observed by direct measurement of the elastic cross
section.

The energy widths of $l=4$ ($g$-wave) Feshbach resonances are
significantly smaller than the average collision energy. As a
consequence, the location of the maximum of the photoassociation
lineshape as a function of magnetic field depends on the
zero-collision-energy resonance field location $B_0$, the
temperature, and the magnetic moment $\mu_{\rm res}$ of the
embedded bound state. In practice, however, we located the
resonances from scattering calculations at $E/k_B=5.3$ $\mu$K and
used the magnetic moment of the $g$-wave Feshbach resonance
obtained from bound-state calculations to extrapolate to zero
collision energy.

\section{Results} \label{sec:results}

Tables \ref{t:el_inel_1}, \ref{t:el_inel_2}, and \ref{t:radiative}
give the magnetic field locations and assignments of the observed
Feshbach resonances. Numbers in parentheses indicate the one
standard deviation uncertainty. The locations of the resonances
are obtained from elastic and inelastic cross-section measurements
or radiative spectroscopy. The theoretical resonance locations are
obtained from coupled-channel scattering calculations at a
$E/k_B$=5.3 $\mu$K collision energy and Born-Oppenheimer
potentials with dispersion coefficients $C_6$=6890 $E_ha_0^6$ and
$C_8$=954600 $E_ha_0^8$, and scattering lengths $a_S=280.3$ $a_0$
and $a_T=2405$ $a_0$ \cite{leo00}. Here $k_B$ is the Boltzmann
constant, $1\,E_h=4.35974$ aJ is a Hartree, and $1\,a_0=0.0529177$
nm is a Bohr. $k_B$5.3 $\mu$K is the average collision energy for
a Cs gas at $T$=3.5 $\mu$K. For resonances observed in the elastic
(inelastic) rates the minimum (maximum) of the line is quoted. The
only exception is the resonance at 7.66 mT, where the position of
the minimum inelastic rate is given. For the resonances observed
by the radiative spectroscopy the maximum of the loss rate and the
maximum of the theoretical inelastic collision rate in the absence
of laser light are presented.

It should be noted that the magnetic field values for the spectral
features presented here should not be confused with the location
of the molecular Feshbach state introduced in
Fig.~\ref{processes}. Typically, the molecular state can be
defined anywhere within the width of the resonance.

\begin{table}
\caption{Location and assignment of the observed Feshbach
resonances. The first three columns define and give the results of
the experiment, where the first column denotes the initial
collision state, the second column indicates elastic (el.),
inelastic (inel.), or radiative (rad.) measurements, and the third
column is the experimental resonance location $B_{exp}$. The next
four columns describe the quantum labels of the resonances. These
are the partial wave of the initial collision state, the partial
wave of the Feshbach state, the molecular spin $f$, and its
projection $m$ obtained from multichannel {\it bound state}
calculations. The last column gives the theoretical resonance
location $B_{th}$ obtained from a multichannel {\it scattering}
calculation at a collision energy of $E/k_B$=5.3 $\mu$K. The
resonance at $B_{ex}=20.66$~mT was not observed in the theoretical
scattering calculation due to its extreme narrowness but
nevertheless could be assigned from bound state calculations.
Several Feshbach resonances have ambiguous assignments (see Text).
} \label{t:el_inel_1}
\begin{tabular}{ccr@{.}lccccr@{.}l}
                                &
 \multicolumn{3}{c}{Experiment} &
 \multicolumn{5}{c}{Theory}    \\
 \cline{2-4}\cline{5-10}
 \multicolumn{4}{c}{}           &
 Inc.                           &
 \multicolumn{3}{c}{Assignment} &
 \multicolumn{2}{c}{}           \\
 \cline{6-8}
 \raisebox{1.5ex}[0pt]{State}   &
 \raisebox{1.5ex}[0pt]{Method}  &
 \multicolumn{2}{c}{\raisebox{1.5ex}[0pt]{$B_{ex}$(mT)}
                            }   &
 Wave                           &
 $l$                            &
 $f$                            &
 $m_f$                          &
 \multicolumn{2}{c}{\raisebox{1.5ex}[0pt]{
         $B_{th}$(mT)}}       \\
\hline
(3,3)+(3,3)   & el.   &  $^{\gamma}$1 & 706(3) & s & s & 6 & 6 &  1 & 70(2)\\
(3,3)+(3,3)   & el.   &  $^{\gamma}$4 & 802(3) & s & d & 4 & 4 &  4 & 79(2)\\
\\
(3,3)+(3,2)   & el.   &  $^{\gamma}$5 & 69(2)  & s & d & 4 & 4  &  5 & 70(2)\\
\\
(4,-4)+(4,-4) & inel. & 10 & 590(3) & s & d & 8 & -6  & 10 & 58(2)\\
(4,-4)+(4,-4) & inel. & 20 & 503(3) & s & d & 8 & -7  &  20 & 49(2)\\
\\
(4,-4)+(4,-3) & inel. & 20 & 66(1)  & s & d & 8 & -6  &  \\
\\
(3,-3)+(3,-3) & inel. &  3 & 005(5) & s & d & 6 & $\approx$-6 &  2 & 99(2)\\
(3,-3)+(3,-3) & inel. &  3 & 305(5) & s & d & 6 & -4 &  3 & 28(2)\\
(3,-3)+(3,-3) & inel. &  8 & 69(2)  & s & d & 8 & -8 &  $^{\beta}$8 & 80(2)\\
(3,-3)+(3,-3) & inel. & 10 & 11(2)  & s & d & 8 & -7 & 10 & 15(2)\\
(3,-3)+(3,-3) & inel. & 10 & 88(2)  & s & d & 8 & -6 & 10 & 90(2)\\
(3,-3)+(3,-3) & inel. & 11 & 81(2)  & s & d & 8 & -5 & 11 & 85(2)\\
(3,-3)+(3,-3) & inel. & 13 & 31(2)  & s & d & 8 & -4 & 13 & 35(2)\\
\\
(3,3)+(4,2)   & inel. &  6 & 17(2)  & s & d & 7 & 5 &  6 & 21(2)\\
(3,3)+(4,2)   & inel. &  $^{\delta}$7 & 66(2)  & s & s & 6 & 5 & 7 & 53(2)\\
(3,3)+(4,2)   & inel. &  8 & 05(2)  & p & p & 6 & 5 &  8 & 08(2)\\
(3,3)+(4,2)   & inel. &  8 & 38(2)  & s & d & 7 & 6 &  8 & 43(2)\\
(3,3)+(4,2)   & inel. & 11 & 00(3)  & s & d & 7 & 5,7  & 11 & 02(2)\\
(3,3)+(4,2)   & inel. & 11 & 20(3)  & s & d & 7 & 5,7  & 11 & 20(2)\\
(3,3)+(4,2)   & inel. & 16 & 22(4)  & s & s & 5 & 5 & 16 & 23(2)\\
(3,3)+(4,2)   & inel. & 18 & 29(5)  & p & p & 5 & 5 & 18 & 45(2)\\
\\
(3,3)+(4,3)   & inel. & 12 & 90(3)  & s & d & 7 & 6 & 12 & 96(2)\\
(3,3)+(4,3)   & inel. & 17 & 30(4)  & p & p & 6 & 6 & 17 & 45(2)\\
(3,3)+(4,3)   & inel. & 22 & 73(5)  & s & d & 7 & 5,7 & 22 & 75(2)\\
(3,3)+(4,3)   & inel. & 23 & 05(5)  & s & d & 7 & 5,7 & 23 & 13(2)\\
\hline \multicolumn{10}{l}{$^{\beta}$ Temperature-dependent
resonance
occurred at $Re(a)=0$.}\\
\multicolumn{10}{l}{$^{\gamma}$ Minimum evaporation rate.}\\
\multicolumn{10}{l}{$^{\delta}$ Minimum inelastic loss.}\\
\end{tabular}

\end{table}

\begin{table}
\caption{Feshbach resonances between two $F=3$ atoms observed in
measurements of inelastic rate coefficients. Columns defined as in
Table~{\protect\ref{t:el_inel_1}}. One bound state could not be
assigned while three resonances have not been observed
experimentally.} \label{t:el_inel_2}
\begin{tabular}{lcr@{.}lccccr@{.}l}
                                &
 \multicolumn{3}{c}{Experiment} &
 \multicolumn{5}{c}{Theory}    \\
 \cline{2-4}\cline{5-10}
 \multicolumn{4}{c}{}           &
 Inc.                           &
 \multicolumn{3}{c}{Assignment} &
 \multicolumn{2}{c}{}           \\
 \cline{6-8}
\multicolumn{1}{c}{\raisebox{1.5ex}[0pt]{State}}   &
 \raisebox{1.5ex}[0pt]{Method}  &
 \multicolumn{2}{c}{\raisebox{1.5ex}[0pt]{$B_{ex}$(mT)}
                            }   &
 Wave                           &
 $l$                            &
 $f$                            &
 $m_f$                          &
 \multicolumn{2}{c}{\raisebox{1.5ex}[0pt]{
         $B_{th}$(mT)}}       \\
\hline
(3,-3)+(3,-3)  & inel. &  2 & 18(2) &   &   &   &     &  $^{*}$ & \\
\\
(3,-3)+(3,-2) & inel. &  3 & 57(2) & s & d & 6 & -5  &  3 & 55(5)\\
(3,-3)+(3,-2) & inel. & 10 & 50(1) & p & f & 7 & -7 & 10 & 50(5)\\
(3,-3)+(3,-2) & inel. & 11 & 04(2) & p & f & 7 & -7 & 11 & 03(5)\\
(3,-3)+(3,-2) & inel. & 11 & 39(2) & p & f & 7 & -7 & 11 & 35(5)\\
(3,-3)+(3,-2) & inel. & 12 & 01(2) & s & d & 8 & -7 & 12 & 03(5)\\
(3,-3)+(3,-2) & inel. & 13 & 01(2) & s & d & 8 & -6 & 13 & 05(5)\\
(3,-3)+(3,-2) & inel. & 14 & 58(2) & s & d & 8 & -5 & 14 & 65(5)\\
(3,-3)+(3,-2) & inel. & 17 & 02(2) & s & d & 8 & -4 & 17 & 05(5)\\
(3,-3)+(3,-2) & inel. &\multicolumn{2}{l}{}& s & d & 8 & -3 & $^{?}$20 & 88(5)\\
\\
(3,-2)+(3,-2) & inel. & $^\tau$14 & 82(2) & s & d & 8 & -7 & 14 & 80(5)\\
(3,-2)+(3,-2) & inel. & $^\tau$16 & 58(2) & s & d & 8 & -6  & 16 & 50(5)\\
(3,-2)+(3,-2) & inel. & $^\tau$19 & 25(2) & s & d & 8 & -5  & 19 & 25(5)\\
(3,-2)+(3,-2) & inel. &\multicolumn{2}{l}{}& s & d & 8 & -4 & $^{?}$23 & 75(5)\\
\\
(3,-3+3,-1) & inel. & 12 & 92(2) & s & d & 8 & -8 & 12 & 90(5)\\
(3,-3+3,-1) & inel. &\multicolumn{2}{l}{}& s & d & 8 & -4 & $^{?}$21 & 15(5)\\
\hline
\multicolumn{10}{l}{$^{*}$ not predicted by calculation.}\\
\multicolumn{10}{l}{$^{?}$ not observed experimentally.}\\
\multicolumn{10}{l}{$^\tau$ has equal contribution from (3,-3)+(3,-1) collision.}\\
\end{tabular}

\end{table}

\begin{table}
\caption{Feshbach resonances and a single shape resonance observed
with radiative spectroscopy. Columns defined as in
Table~{\protect\ref{t:el_inel_1}}. Several Feshbach resonances
have ambiguous assignments (see Text).} \label{t:radiative}
\begin{tabular}{ccr@{.}lccccr@{.}l}
                                &
 \multicolumn{3}{c}{Experiment} &
 \multicolumn{5}{c}{Theory}    \\
 \cline{2-4}\cline{5-10}
 \multicolumn{4}{c}{}           &
 Inc.                           &
 \multicolumn{3}{c}{Assignment} &
 \multicolumn{2}{c}{}           \\
 \cline{6-8}
 \raisebox{1.5ex}[0pt]{State}   &
 \raisebox{1.5ex}[0pt]{Method}  &
 \multicolumn{2}{c}{\raisebox{1.5ex}[0pt]{$B_{ex}$(mT)}
                            }   &
 Wave                           &
 $l$                            &
 $f$                            &
 $m_f$                          &
 \multicolumn{2}{c}{\raisebox{1.5ex}[0pt]{
         $B_{th}$(mT)}}       \\
\hline
(3,3)+(3,3) & rad. &  1 & 102(3) & s & g & 4 & 2 & 1 & 12(2)\\
(3,3)+(3,3) & rad. &  1 & 437(3) & s & g & 4 & 3 & 1 & 46(2)\\
(3,3)+(3,3) & rad. &  1 & 506(3) & s & g & 6 & 5 & 1 & 51(2)\\
(3,3)+(3,3) & rad. &  1 & 83(1)  & s & d &   &   & $^{\epsilon}$1 & 86(2)\\
(3,3)+(3,3) & rad. &  1 & 990(3) & s & g & 4 & 4 & 2 & 01(2)\\
(3,3)+(3,3) & rad. &  4 & 797(3) & s & d & 4 & 4 & 4 & 77(2)\\
(3,3)+(3,3) & rad. &  5 & 350(3) & s & g & 2 & 2 & 5 & 43(2)\\
(3,3)+(3,3) & rad. & 11 & 278(3) & s & d & 6 & 4 & 11 & 32(2)\\
(3,3)+(3,3) & rad. & 13 & 106(3) & s & d & 4 & 4 & 13 & 19(2)\\
\\
(3,3)+(3,2) & rad. &  0 & 78(1) & s & g & 6 & 3 & 0 & 83(2)\\
(3,3)+(3,2) & rad. &  1 & 13(1) & s & g & 4 & 1 & 1 & 17(2)\\
(3,3)+(3,2) & rad. &  1 & 47(1) & s & g & \multicolumn{2}{c}{4 2 or 6 4} & 1 & 54(2)\\
(3,3)+(3,2) & rad. &  1 & 66(1) & p & f & 3 & 2 & 1 & 64(2)\\
(3,3)+(3,2) & rad. &  2 & 09(1) & s & g & 4 & 3 & 2 & 16(2)\\
(3,3)+(3,2) & rad. &  2 & 21(1) & p & f & \multicolumn{2}{c}{1 1 or 3 3}   & 2 & 18(2)\\
(3,3)+(3,2) & rad. &  2 & 36(1) & p & f & \multicolumn{2}{c}{1 1 or 3 3}   & 2 & 33(2)\\
(3,3)+(3,2) & rad. &  3 & 60(1) & s & g & 4 & 4 & 3 & 70(2)\\
(3,3)+(3,2) & rad. &  3 & 81(1) & p & f & 5 & 1 & 3 & 78(2)\\
(3,3)+(3,2) & rad. &  4 & 68(1) & p & \multicolumn{3}{c}{f 5 2 or p $f$=5}  & 4 & 70(2)\\
(3,3)+(3,2) & rad. &  4 & 93(1) & p & \multicolumn{3}{c}{f 5 2 or p $f$=5}  & 4 & 89(2)\\
(3,3)+(3,2) & rad. &  4 & 99(1) & p & \multicolumn{3}{c}{f 5 2 or p $f$=5}  & 4 & 99(2)\\
(3,3)+(3,2) & rad. &  5 & 70(1) & s & d & 4 & 4 & 5 & 70(2)\\
(3,3)+(3,2) & rad. &  5 & 77(1) & p & p & 5 & 3,4,5  & 5 & 78(2)\\
(3,3)+(3,2) & rad. &  5 & 87(1) & p & p & 5 & 3,4,5  & 5 & 86(2)\\
(3,3)+(3,2) & rad. &  5 & 97(1) & p & p & 5 & 3,4,5  & 5 & 98(2)\\
\hline
\multicolumn{10}{l}{$^{\epsilon}$ $d$-wave shape resonance}\\
\end{tabular}

\end{table}

The theoretical uncertainties are a consequence of our limited
ability to model the experiments, and are obtained by combining
the uncertainties due to the spread in collision energies of a
thermal Cs gas with the magnetic field dependence of the Feshbach
resonance. Any discrepancy between theory and experiment in the
tables that lies outside the error bars indicate deficiencies in
the shape of the two Born-Oppenheimer potentials.

The assignment of the resonances is obtained by combining
information from scattering and bound state calculations. The
initial collision partners that lead to a resonance are determined
from theoretical scattering calculations, and experimentally by
varying the relative population of hyperfine states in the Cs gas
and comparing the relative strength of the resonances. The
incoming partial wave is obtained from scattering calculations.
The assignment of quantum numbers $l$, $f$, and $m_f$ is made on
the basis of bound-state calculations. One resonance could not be
assigned.

Levels with the same $l$,$f$,$m_f$ symmetry but different
$M=m_l+m_f$ are degenerate except for small splittings from the
second-order spin-orbit and spin-spin interactions.  For even
partial waves this is not an issue as losses from the $s$-wave
entrance channel at collision energies of the order of a few
microkelvin are much larger than those from $d$-wave entrance
channels and thus only resonances with $M$ equal to the sum of the
magnetic quantum number of the initial hyperfine states,
$M=m_{fa}+m_{fb}$, are observed. For collisions between atoms in
unlike hyperfine states contributions from $p$-wave collisions
cannot be ignored and nearly degenerate Feshbach resonances from
three $M$ values appear.

A good example of the complexity for odd $l$ resonances can be
found between 4.0 mT and 6.0 mT in the radiative Feshbach spectrum
from a $(3,3)+(3,2)$ collision.  In this region six nearly
degenerate resonances are labeled $l=f$, $f=5$, $m_f=2$ or $l=p$,
$f=5$. The $m_f$ labels of the $p$-wave resonance could not be
assigned. It turns out that this resonance has a small magnetic
moment and at $B\approx$ 5 mT the different $m_f$ components are
nearly degenerate.  Mixing between different $m_f$ components for
the same $M$ due to second-order spin-orbit and spin-spin
interactions leads to shifts that are comparable to the spacings
due to the Zeeman interaction. The $l=f$ resonances, which have a
larger magnetic moment, accidentally reside in the same magnetic
field region.

We have assigned more than one set of labels to the resonances
between 4.0 mT and 6.0 mT, because $m_f$ and to a lesser degree
$l$ and $f$ are not good quantum labels.  There are eight
different  $l=p$, $f=5$ and $l=f$, $f=5$, $m_f=2$ resonances
between 4.0 mT and 6.0 mT. Six of these eight resonances are due
to the $p$-wave symmetry. This can easily be checked by noting
that only $M$=4, 5, and 6 can lead to these odd $l$ resonances.
Not all eight resonances have been seen which might be due to the
fact some are not resolved or unobserved by radiative
spectroscopy. We did not perform quantitative {\it bound state}
calculations as too many channels must be coupled together. The
theoretical field locations listed in the last column of the
tables have been obtained from scattering calculations where all
states are included at the cost of losing the ability to assign
quantum labels.

\begin{figure}
\begin{turn}{-90}
\includegraphics[width=2.4in]{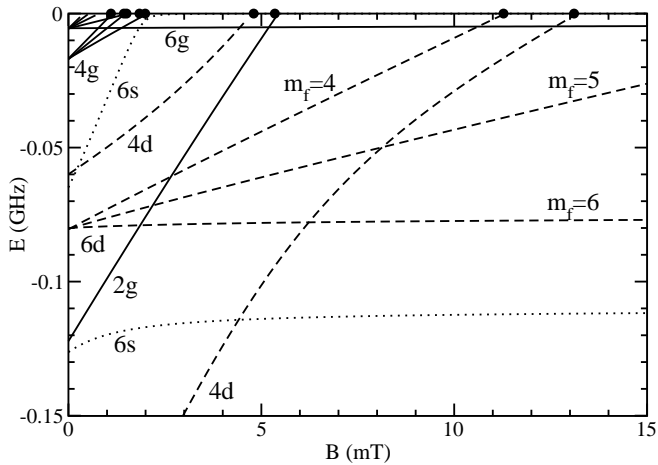}
\end{turn}
\caption{Total angular momentum projection $M$=6, $s$, $d$, and
$g$-wave bound state energies as a function of magnetic field. The
zero of energy corresponds to the $(F=3,m_F=3)+(F=3,m_F=3)$
dissociation limit. Dotted, dashed, and solid lines correspond to
$l$=$s$,$d$, and $g$ states respectively. Furthermore, each curve
is labeled by the quantum numbers $fl$, i.e. $fl=6s$. The
molecular spin $f$ is a zero field quantum number. For the $fl=6d$
states the magnetic quantum number $m_f$ is also indicated. The
filled circles represent the observed threshold resonances.}
\label{longM6}
\end{figure}

Fig.~\ref{longM6} shows even $l$ $M$=6 bound states below the
lowest molecular hyperfine state $(3,3)+(3,3)$ as a function of
magnetic field.  Each bound state is labeled with $l$, $f$, and
$m_f$. A resonance occurs when a bound state crosses zero energy.
The frequency range shown in the figure is sufficient for the
assignment of all $B<$15 mT $s$, $d$, and $g$-wave Feshbach
resonances in the collision between $|3,3\rangle$ Cs atoms. The
filled circles mark the observed threshold resonances in a Cs gas
at $T$=3.5 $\mu$K ($k_BT/h\sim 0.1$ MHz). Agreement between theory
and experiment is sufficiently good that assignments can be made
even though discrepancies exist. These discrepancies are caused by
the (slightly) incorrect shape of the Born-Oppenheimer potentials
and the approximations in the bound state calculations.  For
fields smaller than 1 mT theory predicts the existence of
additional resonances.

The number of coupled channels for $M$=6 and $l$=s, d, and g is
74. However, as discussed in the previous section, for nonzero
applied magnetic field the coupling between different partial
waves and $m_f$'s is due to weak second-order spin-orbit and
spin-spin interactions. Consequently, for most resonances in
Fig.~\ref{longM6} assignment is unambiguous using independent
bound-state calculations that only include states of a given $l$
and $m_f$.  In fact, the curves in the figure have been obtained
in this way.  However, it should be realized that crossings
between bound states shown in the figure are actually avoided when
second-order spin-orbit and spin-spin interactions are included.
At zero magnetic field coupling between channels with different
molecular spin $f$ is also small.  The assignment of $f$ is
obtained by retracing a bound state to zero magnetic field and
noting that $|m_f|\leq f$.

\begin{figure}
\begin{turn}{-90}
\includegraphics[width=2.5in]{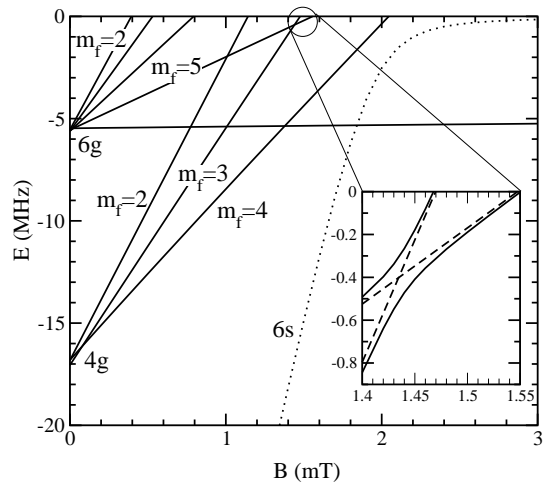}
\end{turn}
\caption{Expanded view of the $M$=6, $s$, $d$, and $g$-wave bound
states shown in Fig.~{\protect\ref{longM6}}. An avoided crossing
between $g$-wave $m_f$=3 and 5 bound state occurs around $B$= 1.4
mT.}
\label{blowupM6}
\end{figure}

A close look at Fig.~\ref{longM6} shows that the lines can roughly
be divided into those that are noticeably curved and those that
appear straight.  A good example of curved lines are the two
$fl=4d$ bound states, while the $6d$ and $4g$ bound states are
good examples of bound states that have a linear magnetic field
dependence. The curved lines are due to broad avoided crossings
that appear because the same $fl$ labeled states readily mix when
a magnetic field is applied. Mixing is due to the interplay of
hyperfine, Born-Oppenheimer, and Zeeman interactions and is
significantly larger than in avoided crossings mediated by
second-order spin-orbit and spin-spin interactions.

The most weakly-bound $fl=6s$ state is bound by about 65 MHz at
zero magnetic field, rises rapidly until it turns over near 2 mT,
and then continues just below the dissociation limit. This avoided
crossing is also shown in Fig.~\ref{blowupM6}.  The bound state
does not run parallel to the dissociation limit.  It becomes a
Feshbach resonance near $B$=50 mT.  The behavior of this bound
state has direct consequences for the $s$-wave scattering length
of two $|f_a m_{a}\rangle=|3,3\rangle$ atoms. Below 1.7 mT  the
scattering length is negative and above this field value it is
positive.  This zero of the scattering length has been observed in
Ref.~\cite{Vul99} and has been used to optimize the
Born-Oppenheimer potentials in Refs~\cite{chi00,leo00}, as well as
in this paper. It is interesting to realize that, as discussed in
Ref.~\cite{leo00}, for $B< 1.7$ mT $d$-wave channels affect the
elastic scattering and must be included in order to obtain an
accurate scattering length.

Some of the $M$=6 and $l$=s, d and g Feshbach resonances below
$B$=3 mT could at first not be assigned from calculations using
states with the same $l$ and $m_f$. Resonances of different $lm_f$
symmetry lie in the same magnetic field region.
Fig.~\ref{blowupM6} shows a blow up of the 0 mT to 3 mT magnetic
field range. The $f$=4 $m_f$=3 and $f$=6 $m_f$=5 $g$-wave bound
states cross just below the dissociation limit and weak couplings
might shift the corresponding Feshbach resonances. The inset shows
the avoided crossing between these $f$=4 $m_f$=3 and $f$=6 $m_f$=5
$g$-wave bound state when the weak coupling between the two bound
states is included.  From the figure it is clear that the avoided
crossing has little influence on the location of the Feshbach
resonances and an $fm_f$ label for each resonance can still be
assigned.

The Born-Oppenheimer potentials that have been used for the
theoretical resonance locations quoted in
Tables~\ref{t:el_inel_1}, \ref{t:el_inel_2}, and \ref{t:radiative}
and Figs.~\ref{longM6} and \ref{blowupM6} are based on the fit in
our previous work~\cite{leo00}, where the major uncertainties in
the calculation of resonance positions arise from the poorly
constrained $C_8$ coefficient. For this Paper we have improved the
Born-Oppenheimer potentials by optimizing the $C_8$ dispersion
coefficient in addition to the $C_6$ coefficient, the strength of
the second-order spin-orbit interaction $S_C$, and the singlet and
triplet scattering lengths \cite{leo00}.  For a given $C_8$ the
potentials are optimized to fit the minima in the elastic
scattering rate of the $(3,3)+(3,3)$ scattering at 1.7064 mT and
4.8017 mT, the $f=6$ $d$-wave resonance in $(3,-3)+(3,-3)$
scattering at 3.0051 mT, and the resonance in $(4,-4)+(4,-4)$
scattering at 20.5029 mT. Over a 10\% range of the $C_8$
coefficient near 900000 $E_ha_0^8$ a linear relationship between
$C_6$, $S_C$, $a_S$, and $a_T$ exists.

\begin{figure}
\begin{turn}{-90}
\includegraphics[width=2.5in]{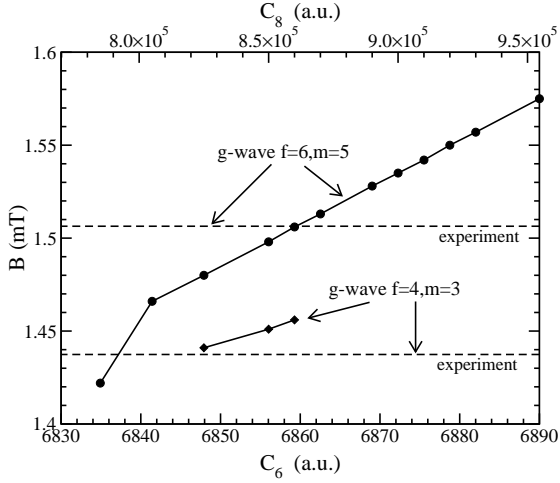}
\end{turn}
\caption{Position of two $M=6$ $g$-wave resonances as a function
of $C_8$ or equivalently $C_6$ for a zero-energy collision of two
Cs atoms in the lowest hyperfine state. The $C_6$ and $C_8$ are
expressed in units of $E_ha_0^6$ and $E_ha_0^8$, respectively. The
dotted lines correspond to the peak of the experimental radiative
Feshbach spectroscopy signal.} \label{C8dep}
\end{figure}

\begin{figure}
\includegraphics[width=3in]{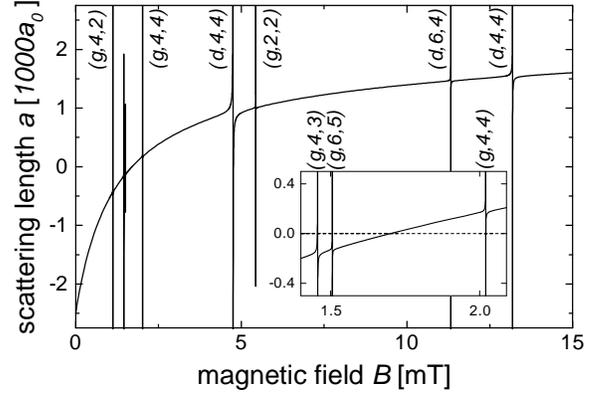}
\caption{Scattering length in the $(F=3, m_F=3)+(F=3, m_F=3)$
scattering channel. Resonances resulted from the $M$=6, $s$, $d$,
and $g$-wave bound states are assigned with quantum number
$(l,f,m_f)$, where $l$ is the orbital angular momentum, $f$ is the
total internal angular momentum and $m_f$ is the magnetic quantum
number. Calculation are done with a magnetic field grid size of
$50\mu$T for off-resonance regions and $100$nT near the narrow
resonances. Inset shows the detail resonance structure near $17$G}
\label{aa}
\end{figure}

\begin{figure}
\includegraphics[width=3in]{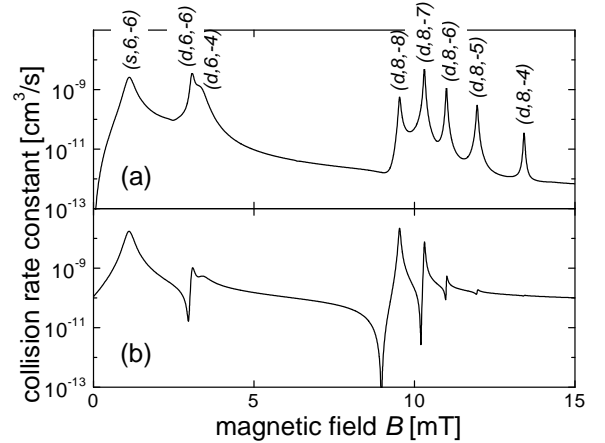}
\caption{Collision rate constants in the $(F=3, m_F=-3)+(F=3,
m_F=-3)$ scattering channel at $E/k_B=1n$K. Inelastic collision
rates $(a)$ and elastic collision rates $(b)$ are calculated with
a magnetic field grid size of $10\mu$T for off-resonance regions
and $1\mu$T near the resonances. Resonances from the $M$=-6, $s$,
$d$, and $g$-wave bound states are included and assigned with
quantum number $(l,f,m_f)$, where the notation is the same as in
Fig.~\ref{aa}.} \label{gg}
\end{figure}

An improved cesium-dimer Hamiltonian is created by fitting to
selected $(3,3)+(3,3)$ $g$-wave resonances. Fig. \ref{C8dep} shows
the $M$=6, $f$=6 $m_f$=5 and $f$=4 $m_f$=3, $g$-wave resonance
position as a function of $C_8$ or equivalently $C_6$. The figure
shows theoretical zero-collision-energy resonance locations
derived from $E/k_B$=5.3 $\mu$K calculations and the magnetic
moments of the resonances. The peak radiative detection signal as
a function of magnetic field is due to collisions at zero energy.
The magnetic moment of the resonances is 170 $\mu$K/mT for the
$f$=6, $m_f$=5 and 550 $\mu$K/mT for the $f$=4, $m_f$=3 state.

The location of the Feshbach resonance found from a bound state
calculation, including only $g$-wave $m_f$=3 and 5 channels, and a
scattering calculation at zero collision energy, which includes
{\it all} $s$-, $d$-, and $g$-wave channels, disagree by about
0.02 mT. This discrepancy is likely due to the limited number of
channels in the bound state calculations.  The magnetic moments,
however, are not expected to be significantly modified.

\begin{table}
\caption{Properties of the singlet X$^1\Sigma_g^+$ and triplet
a$^3\Sigma_u^+$ Born-Oppenheimer potentials and the second-order
spin-orbit interaction that give the best fit to all data on
collisions between ultra-cold Cs atoms. The $C_6$ and $C_8$ are
expressed in units of $E_ha_0^6$ and $E_ha_0^8$, respectively. The
singlet, $a_S$, and triplet, $a_T$, scattering lengths are in
units of $a_0$.  $S_C$ is dimensionless. One-standard deviation
uncertainties are given.} \label{t:final}

\begin{tabular}{c|cc}
           &   value   & uncertainty $(\%)$ \\
\hline
   $C_6$   &   6860    &  0.36       \\
   $C_8$   &  860\,000   & 8.7 \\
   $a_S$   &   280.37   & 0.02  \\
   $a_T$   &  2440      & 1.0    \\
   $S_C$   &  2.6    & 19
\end{tabular}
\end{table}

Table \ref{t:final} summarizes our best fit. Based on the
collision parameters, cold collision properties of cesium atoms
can be readily calculated in various scattering channels. In
particular, $(3,3)+(3,3)$ and $(3,-3)+(3,-3)$ scattering length is
$-$2510 $a_0$ at zero magnetic field. In the presence of the
magnetic field, Fig.~\ref{aa} shows the scattering length of the
$(3,3)+(3,3)$ channel as a function of magnetic field;
Fig.~\ref{gg} shows the collision rate constants in the
$(3,-3)+(3,-3)$ channel at the collision energy of $E=k_B$ 1nK.
All $s$, $d$, and $g$ wave channels are included in the
calculation. These two states are particularly interesting in the
experiments of cesium Bose-Einstein condensation \cite{csbec}. The
resonances are identified and labelled by the quantum numbers of
the associated molecular states.

\section{Conclusion}

We have measured $>60$ magnetic-field-induced Feshbach resonances
in the collision of ultracold ground state cesium atoms. Of all
the alkali-metal species cesium is shown to have the richest
resonance structure.

The field resonances have been observed in elastic collision
rates, evaporation loss rates, collision relaxation rates, as well
as in radiative collision resonance experiments (radiative
Feshbach spectroscopy). The last experiments have been
instrumental in observing $l=4$ $g$-wave Feshbach resonances with
$\sim$mG resonance width.

Based on the previous work \cite{leo00}, we have also improved the
model for the Cs-Cs collision and in addition used multi-channel
bound state calculations to assign each Feshbach resonance with
pertinent quantum numbers. The quantum numbers correlate each
resonance to a molecular bound state at zero magnetic field. We
identify 3 $s$-wave, 6 $p-$wave, 32 $d-$wave, 10 $f-$wave and 10
$g-$wave Feshbach resonances and one shape resonance. One
resonance in the $(3,3)+(3,3)$ channel could not be identified and
is possibly a three-body collision resonance or a two-body
Feshbach resonance with very high partial-waves number.

The model has been used to calculate the molecular energy
structure below the $(F=3)+(F=3)$ continuum and the collision
properties in the $(3,3)+(3,3)$ and $(3,-3)+(3,-3)$ scattering
channels. These data will provide crucial information for
experiments on cesium Bose-Einstein condensation \cite{csbec} and
cesium molecules \cite{csmol}.

In general, this paper presents results of a successful
collaboration of experimental and theoretical efforts and resolves
the collision anomalies of cesium atoms. The excellent agreement
on $>$60 resonances between experiment and theory with only $5$
parameters marks a triumph of the predictive power of atomic
interaction theory. Since the experimental determination of the
resonance locations is better than the theoretical estimate.
Further improvements would require more flexibility in the
(short-range shape) of the two Born-Oppenheimer potentials.
Moreover, theoretical modeling will need to use thermally averaged
elastic and inelastic rates as well as an improved model of the
radiative lineshape.

\section{Acknowledgement}
C.C. would like to thank P.S. Julienne for discussions. This work
was supported in part by grants from the AFOSR and the NSF.


\begin{references}
\bibitem{fesh0} E. Tiesinga, B.J. Verhaar, and H.T.C. Stoof, Phys.\ Rev.\ A {\bf 47}, 4114
(1993)
\bibitem{fesh} S. Inouye, M. Andrews, J. Stenger, H.-J. Miesner, S.
Stamper-Kurn, and W. Ketterle, Nature {\bf 392}, 151 (1998).
\bibitem{implo} E.A. Doney, N.R. Claussen, S.L. Cornish, J.L. Roberts, E.A. Cornell and C.E.
Wieman, Nature {\bf 412}, 295 (2001).
\bibitem{cohr} E.A. Donley, N.R. Claussen, S.T. Thompson, C.E.
Wieman, Nature {\bf 417}, 529 (2002).
\bibitem{sol} L. Khaykovich, F. Schreck, G. Ferrari, T. Bourdel, J. Cubizolles, L.D. Carr, Y. Castin, and C. Salomon, Science {\bf 296}, 1290 (2002); K.E. Strecker, G.B. Partridge, A.G. Truscott, R.G.
Hulet, Nature {\bf 417}, 150 (2002).
\bibitem{chi03} C. Chin, A.J. Kerman, V. Vuleti\'c, S. Chu, Phys.\ Rev.\ Lett.\ {\bf 90}, 033201 (2003).
\bibitem{csmol} J. Herbig, T. Kraemer, M. Mark, T. Weber, C. Chin,
H.-C. N\"{a}gerl and R. Grimm, Science {\bf 301}, 1510 (2003).
\bibitem{mol} C.A. Regal, C. Ticknor, J.L. Bohn, D.S. Jin, Nature {\bf 424}, 47 (2003);
K.E. Strecker, G.B. Partridge, R.G. Hulet, Phys.\ Rev.\ Lett.\
{\bf 91}, 080406 (2003); J. Cubizolles,T. Bourdel, S.J.J.M.F.
Kokkelmans, G.V. Shlyapnikov, C. Salomon, Phys.\ Rev.\ Lett.\ {\bf
91}, 240401 (2003); S. Jochim, M. Bartenstein, A. Altmeyer, G.
Hendl, S. Riedl, C. Chin, J. Hecker Denschlag and R. Grimm, Phys.\
Rev.\ Lett.\ {\bf 91}, 240402 (2003); K. Xu, T. Mukaiyama, J.R.
Abo-Shaeer, J.K. Chin, D. Miller, and W. Ketterle, Phys.\ Rev.\
Lett.\ {\bf 91}, 210402 (2003).
\bibitem{molbec} S. Jochim, M. Bartenstein, A. Altmeyer, G. Hendl, S. Riedl, C. Chin, J. Hecker Denschlag, and R. Grimm, published online 13 November
2003; 10.1126/science.1093280 (\emph{Science Express}); M.
Greiner, C.A. Regal and D.S. Jin, Nature {\bf 426}, 537 (2003);
M.W. Zwierlein, C.A. Stan, C.H. Schunck, S.M.F. Raupach, S. Gupta,
Z. Hadzibabic and W. Ketterle, {\bf 91}, 250401 (2003).
\bibitem{gib93} K. Gibble and S. Chu, Phys.\ Rev.\ Lett.\ {\bf 70}, 1771 (1993).
\bibitem{xsec} M. Arndt, M. Ben Fahan, D. Gu\'{e}ry-odelin, M.W. Reynolds, and D. Dalibard, Phys.\ Rev.\ Lett.\ {\bf 79}, 625 (1997).
\bibitem{becfail} D. Gu\'{e}ry-Odelin, J. Soeding, P. Desbiolles, and Jean
Dalibard, Opt.\ Express\ {\bf 2}, 323 (1998).
\bibitem{wei85} H. Weichenmeier, U. Diemer, M. Wahl, M. Raab, W.
W\"{u}ller, and W. Demtr\"{o}der, J.\ Chem.\ Phys.\ {\bf 82}, 5354
(1985).
\bibitem{clock} P.J. Leo, P.S. Julienne, F.H. Mies, and C.J.
Williams, Phys.\ Rev.\ Lett.\ {\bf 86}, 3743 (2001).
\bibitem{leo00} P.J. Leo, C.J. Williams, P.S. Julienne, Phys.\ Rev.\ Lett.\ {\bf 85}, 2721 (2000).
\bibitem{vul99} V. Vuleti\'{c}, C. Chin, A.J. Kerman and S. Chu, Phys.\ Rev.\ Lett.\ {\bf 83}, 943 (1999).
\bibitem{chi00} C. Chin, V. Vuleti\'c, A.J. Kerman, S. Chu, Phys.\ Rev.\ Lett.\ {\bf 85}, 2717 (2000).
\bibitem{Vul99} V. Vuleti\'c, A.J. Kerman, C. Chin, and S. Chu, Phys. Rev. Lett. {\bf 82}, 1406 (1999).
\bibitem{Heinzen} Ph. Courteille, R.S. Freeland, D.J. Heinzen, F.A. van Abeelen and B.J. Verhaar, Phys.\ Rev.\ Lett.\ {\bf 81}, 69 (1998).
\bibitem{deu98} I.H. Deutsch and P.S. Jessen, Phys.\ Rev.\ A {\bf 57}, 1972 (1998).
\bibitem{vul98} V. Vuleti\'{c}, C. Chin, A.J. Kerman, and S. Chu, Phys.\ Rev.\ Lett.\ {\bf 81}, 5768 (1998).
\bibitem{ker00} A.J. Kerman, V. Vuleti\'{c}, C. Chin, and S. Chu, Phys.\ Rev.\ Lett.\ {\bf 84}, 439 (2000).
\bibitem{fri98} S. Friebel, C. D'Andrea, J. Walz, M. Weitz, and T. W. H\"{a}nsch, Phys.\ Rev.\ A {\bf 57}, R20 (1998).
\bibitem{dep01} M.T. DePue, D.J. Han and D.S. Weiss, Phys.\ Rev.\ A 63, 023405 (2001).
\bibitem{arn97} M. Arndt, M. Ben Fahan, D. Guery-Odelin, M.W. Reynolds, and J. Dalibard, Phys.\ Rev.\ Lett.\ {\bf 79}, 625 (1997).
\bibitem{sod98} J. Soding, D. Guery-Odelin, P. Desbiolles, G. Ferrari, and J. Dalibard, Phys.\ Rev.\ Lett.\ {\bf 80}, 1869 (1998).
\bibitem{gue98} D. Guery-Odelin, J. Soding, P. Desbiolles, and J. Dalibard, Opt.\ Express.\ {\bf 2}, 323 (1998).
\bibitem{tino03} T. Weber, J. Herbig, M. Mark, H.-C. N\"{a}gerl, and R. Grimm, Phys.\ Rev.\ Lett.\ {\bf 91}, 123201 (2003).
\bibitem{rob98} J. L. Roberts, N. R. Claussen, James P. Burke, Jr., Chris H. Greene, E. A. Cornell, and C. E. Wieman, Phys.\ Rev.\ Lett.\ {\bf 81}, 5109 (1998).
\bibitem{Ari77} E. Arimondo, M. Inguscio, and P. Violino, Rev.\ Mod.\ Phys.\ {\bf 49}, 31 (1977).
\bibitem{Gordon}  R.G. Gordon, J.\ Chem.\ Phys.\ {\bf 51}, 14 (1969); R.G. Gordon, Methods Comput. Phys. {\bf 10}, 81 (1971).
\bibitem{DVR} D.T. Colbert and W.H. Miller, J. Chem. Phys. {\bf 96}, 1982 (1992); E. Tiesinga, C.J. Williams, and P.S. Julienne, Phys.\ Rev.\ A {\bf 57}, 4257 (1998).
\bibitem{MQDT} M. Raoult and F.H. Mies, Phys. Rev. A {\bf 62}, 012708 (2000).
\bibitem{Vog98} J.M. Vogels, B.J. Verhaar, and R.H. Blok, Phys. Rev. A {\bf 57}, 4049 (1998).
\bibitem{Laue02}  T. Laue, E. Tiesinga, C. Samuelis, H. Kn\"ockel, E. Tiemann, Phys. Rev. A, {\bf 65}, 023412 (2002).
\bibitem{Nap94} R. Napolitano, J. Weiner, C. J. Williams, and P. S. Julienne, Phys. Rev. Lett. {\bf 73}, 1352 (1994).
\bibitem{csbec} T. Weber, J. Herbig, M. Mark, H.-C. N\"{a}gerl, R.
Grimm, Science {\bf 299}, 232 (2003).

\end{references}
\end{document}